\pdfoutput=1
\documentclass[useAMS,usenatbib]{mn2e}

\usepackage{times}
\usepackage{amssymb,graphicx}
\usepackage[normalem]{ulem}
\usepackage{changebar}

\newcommand{\avg}[1]{\langle#1\,\rangle}  
\newcommand{\cmcube}{\ensuremath{\,\mathrm{cm^{-3}}}} 
\newcommand{\degree}{\ensuremath{^{\circ}}}
\newcommand{\dif}{\,\mathrm{d}} 
\newcommand{\DM}{\ensuremath{\,\mathrm{DM}}} 
\newcommand{\EMp}{\ensuremath{\,\mathrm{EM}}} 
\newcommand{\Fd}{\ensuremath{\,F_\mathrm{d}}}
\newcommand{\Fv}{\ensuremath{\,F_\mathrm{v}}} 
\newcommand{\HI}{\ensuremath{\mathrm{H\,\scriptstyle I}}} 
\newcommand{\HII}{\ensuremath{\mathrm{H\,\scriptstyle II}}} 
\newcommand{\K}{\ensuremath{\,\mathrm{K}}} 
\newcommand{\kms}{\ensuremath{\,\mathrm{km}\,\mathrm{s}^{-1}}}
\newcommand{\kpc}{\ensuremath{\,\mathrm{kpc}}}
\newcommand{\Myr}{\ensuremath{\,\mathrm{Myr}}}
\newcommand{\n}{\ensuremath{\,n_\mathrm{e}}} 
\newcommand{\Nc}{\ensuremath{\,N_\mathrm{c}}} 
\newcommand{\nH}{\ensuremath{\,n_\mathrm{H\,\scriptstyle I}}}
\newcommand{\NH}{\ensuremath{\,N(\mathrm{H\,\scriptstyle I})}}
\newcommand{\pc}{\ensuremath{\,\mathrm{pc}}}
\title[Density PDFs of diffuse gas]{Density probability distribution functions of diffuse gas in the Milky Way}
\author[E. M. Berkhuijsen and A. Fletcher]{E. M. Berkhuijsen$^1$
	\thanks{E-mail: eberkhuijsen@mpifr-bonn.mpg.de} and 
	A. Fletcher$^2$
	\thanks{E-mail: andrew.fletcher@ncl.ac.uk}\\
$^{1}$Max-Planck-Institut f\"ur Radioastronomie, Auf dem H\"ugel 69,
          53121 Bonn, Germany.\\
$^{2}$School of Mathematics and Statistics, Newcastle University, 
          Newcastle upon Tyne, NE1 7RU, U.K.}
\begin{document}

\date{}

\pagerange{\pageref{firstpage}--\pageref{lastpage}} \pubyear{}

\maketitle

\label{firstpage}
%________________________________________________________________  
\begin{abstract}
In a search for the signature of turbulence in the diffuse interstellar medium in gas density distributions, we determined the probability distribution functions (PDFs) of the average volume densities of the diffuse gas. The densities were derived from dispersion measures and $\HI$ column densities towards pulsars and stars at known distances. The PDFs of the average densities of the diffuse ionized gas (DIG) and the diffuse atomic gas are close to lognormal, especially when lines of sight at $|b|<5\degree$ and $|b|\ge5\degree$ are considered separately. The PDF of $\avg{\nH}$ at high $|b|$ is twice as wide as that at low $|b|$. The width of the PDF of the DIG is about $30$ per cent smaller than that of the warm $\HI$ at the same latitudes. The results reported here provide strong support for the existence of a lognormal density PDF in the diffuse ISM, consistent with a turbulent origin of density structure in the diffuse gas.
\end{abstract}


%________________________________________________________________  
\section{Introduction}
Simulations of the interstellar medium (ISM) have shown that, if isothermal turbulence is
shaping the structure of the medium, the density distribution becomes 
lognormal \citep[ and references therein]{Elmegreen:2004}. However, 
\citet{Madsen:2006} found large temperature differences between 
lines of sight through the warm ionized medium, which seem inconsistent 
with an isothermal gas. Also in the MHD simulations of \citet{Avillez:2005} and \citet{Wada:2007} the ISM became non-isothermal.
The latter authors showed that 
for a large enough volume, and for a long enough simulation run, the physical
processes causing the density variations in the ISM in a galactic
disc can be regarded as random and independent events. Therefore,
the PDF of log(density) becomes Gaussian and the density PDF
lognormal, although the medium is not isothermal. The medium is
inhomogeneous on a local scale, but in a quasi-steady state on a
global scale. 

The shape of the gas density PDF can be an important component in theories of star formation. \citet{Elmegreen:2002} showed that a Schmidt-type power-law relation between the star formation rate per unit area and the gas surface density can be deduced if the density PDF is lognormal and star formation occurs above a threshold density: the dispersion of the lognormal density PDF is a key parameter in the model. However, from the existing simulations it is not yet clear whether a lognormal density PDF in galaxies is universal and which factors determine their shape \citep{Wada:2007}.

Little observational evidence exists to test the results of the 
simulations. \citet{Wada:2000} showed that the luminosity function 
of the HI column density in the Large Magellanic Cloud is lognormal.
Recently, \citet{Hill:2007, Hill:2008} derived a lognormal distribution of the
emission measures perpendicular to the Galactic plane of the DIG in 
the Milky Way, observed in the Wisconsin H$\alpha$ Mapper survey \citep{Haffner:2003} above Galactic latitudes of $10\degree$. They note that emission measures towards classical HII 
regions also fit a lognormal distribution, with different parameters.
Furthermore, \citet{Tabatabaei:2008} found a lognormal distribution of
the emission measures derived from an extinction-corrected H-$\alpha$ map 
of the nearby galaxy M33 \citep{Tabatabaei:2007}. \citet{Gaustad:1993} plotted the distribution of the local volume density of dust near stars within 400\pc\ of the Sun, which is also consistent with a lognormal distribution. In this Letter we 
present the PDFs of average volume densities of the DIG and the 
diffuse atomic gas in the Milky Way, and show that they are consistent 
with lognormal distributions as well. 

%________________________________________________________________  
\section{Basics and Data}

We investigated the density PDFs of the DIG 
and of diffuse atomic hydrogen gas ($\HI$) in the solar neighbourhood. 

Various volume densities of the DIG can be obtained from  the dispersion 
measure $\DM$ (in $\cmcube \pc$) and emission measure $\EMp$ 
(in $\mathrm{cm}^{-6} \pc$) towards 
a pulsar at known distance $D$ (in $\pc$) from the relations:
\begin{equation}
\DM = \int_{0}^{D}\n(l)\dif l  = \avg{\n}\, D =  \Nc \Fd\, D,
\label{eq:DM}
\end{equation}
\begin{equation}
\EMp = \int_{0}^{D}\n^2(l)\dif l = \avg{\n^2}\, D = \Nc^2 \Fd\,D,
\label{eq:EM}
\end{equation}
where $\n(l)$ is the local electron density at point $l$
along the line-of-sight (LOS), $\avg{\n}$ and $\avg{\n^2}$ are averages along
$D$ and $\Fd$ is the fraction of the line of sight in clouds of average
density $\Nc$ \citep[see fig. 1 in][]{Berkhuijsen:2006}.
The final equality in Eq.~\ref{eq:EM} is \emph{only} valid when the average density of \emph{every} cloud $n_c$ along the LOS is the same: then $\avg{\n^2}=\avg{n_c^2}\Fd= \Nc^2\Fd$. Thus $\Nc$ and $\Fd$ are crude approximations of the true average cloud density and filling factor along a LOS, but given the large number of different LOS in our sample their mean and dispersion are reasonable estimators of $\Nc$ and$\Fd$.

Combining Equations~(\ref{eq:DM}) and (\ref{eq:EM}) we have\footnote{Note that we write $\Fd$, $\Fv$ and $\Nc$ where \citet{Berkhuijsen:2006} used $\bar{f_\mathrm{d}}$, $\bar{f_\mathrm{v}}$ and $\bar{n_\mathrm{c}}$.}
\begin{equation}
\Nc  = \frac{\EMp}{\DM},
\label{eq:Nc}
\end{equation}
and 
\begin{equation}
\Fd = \frac{\DM^2}{\EMp \ D} = \frac{\avg{\n}}{\Nc}.
\label{eq:Fd}
\end{equation}
The line-of-sight filling factor $\Fd$ approximates the volume
filling factor $\Fv$ if there are several clouds along the line of
sight. As this will generally be the case, we take $\Fv=\Fd$.

We used the densities derived from two pulsar samples that were 
originally selected for studies of the volume filling factor of
the DIG.
\begin{enumerate}
\item 34 pulsars at observed distances known to better than $50$ per cent, collected by \citet{Berkhuijsen:2008}. The pulsar distances are in the range $0.1<D<9.5\kpc$, with a mean distance of $2.4\kpc$ and a standard deviation of $2.9\kpc$ (the spread is large because 21 pulsars lie at $D<2\kpc$). 
\item 157 pulsars with distances obtained from DM and the model of the distribution of free electrons in the MW of \citet{Cordes:2002}, collected by \citet{Berkhuijsen:2006}. These pulsars lie in the range $0.1<D<6\kpc$, with mean distance $1.7\kpc$ and standard deviation $1.0\kpc$.
\end{enumerate}
Apart from 6 pulsars in the small sample, all pulsars are located
at $|b|\ge5\degree$ in order to ensure that the lines of sight towards the 
pulsars probe the DIG and not denser $\HII$ regions. The dispersion
measures were taken from the catalogue of \citet{Manchester:2005}.
The emission measures in the direction of the pulsars were obtained
from H$\alpha$ surveys \citep{Haffner:2003, Finkbeiner:2002} and 
corrected for extinction \citep{Diplas:1994b, Dickinson:2003} as well as for H$\alpha$ emission originating beyond the pulsars. 
We refer to the work of \citet[][ hereafter called BMM]{Berkhuijsen:2006} 
and \citet{Berkhuijsen:2008} for further details.\footnote{The data used in our analysis are available, on request, from EMB.}

\citet{Diplas:1994b} studied the scale height of the diffuse
dust and the diffuse atomic gas using 393 stars in the Galaxy. We 
calculated the average $\HI$ volume density, $\avg{\nH}$ (in $\cmcube$), from 
the column density $\NH$ (in $\mathrm{cm}^{-2}$), corrected for contributions 
from the star, and the distance to the star as given in Table 1 
of \citet{Diplas:1994a}:
\begin{equation}
\NH =  \int_{0}^{D}\nH(l)\dif l  =  \avg{\nH}\, D,
\label{eq:NH}
\end{equation}
where $\nH(l)$ (in $\cmcube$) is the local volume density at distance $l$ 
along the line of sight. We removed 18 stars from the sample with 
denser clouds in their lines of sight indicated in Fig. 9 of \citet{Diplas:1994b}, leaving the data towards 375 stars for analysis. The stars in this sample have distances in the range $0.1<D<11\kpc$, with a mean distance of $2.2\kpc$ and standard deviation $1.7\kpc$. \footnote{We do not consider the PDFs of the column densities $\DM$, $\EMp$ and $\NH$, because they are influenced by the distributions of the distances to the pulsars and stars in the samples.}

%________________________________________________________________  
\section{Results}
\label{sec:results}

\subsection{Density PDFs of the DIG}
\label{subsec:results:DIG}

%% figure
\begin{figure}
\centering
\begin{minipage}[c]{0.235\textwidth}
\includegraphics[width=1.0\textwidth]{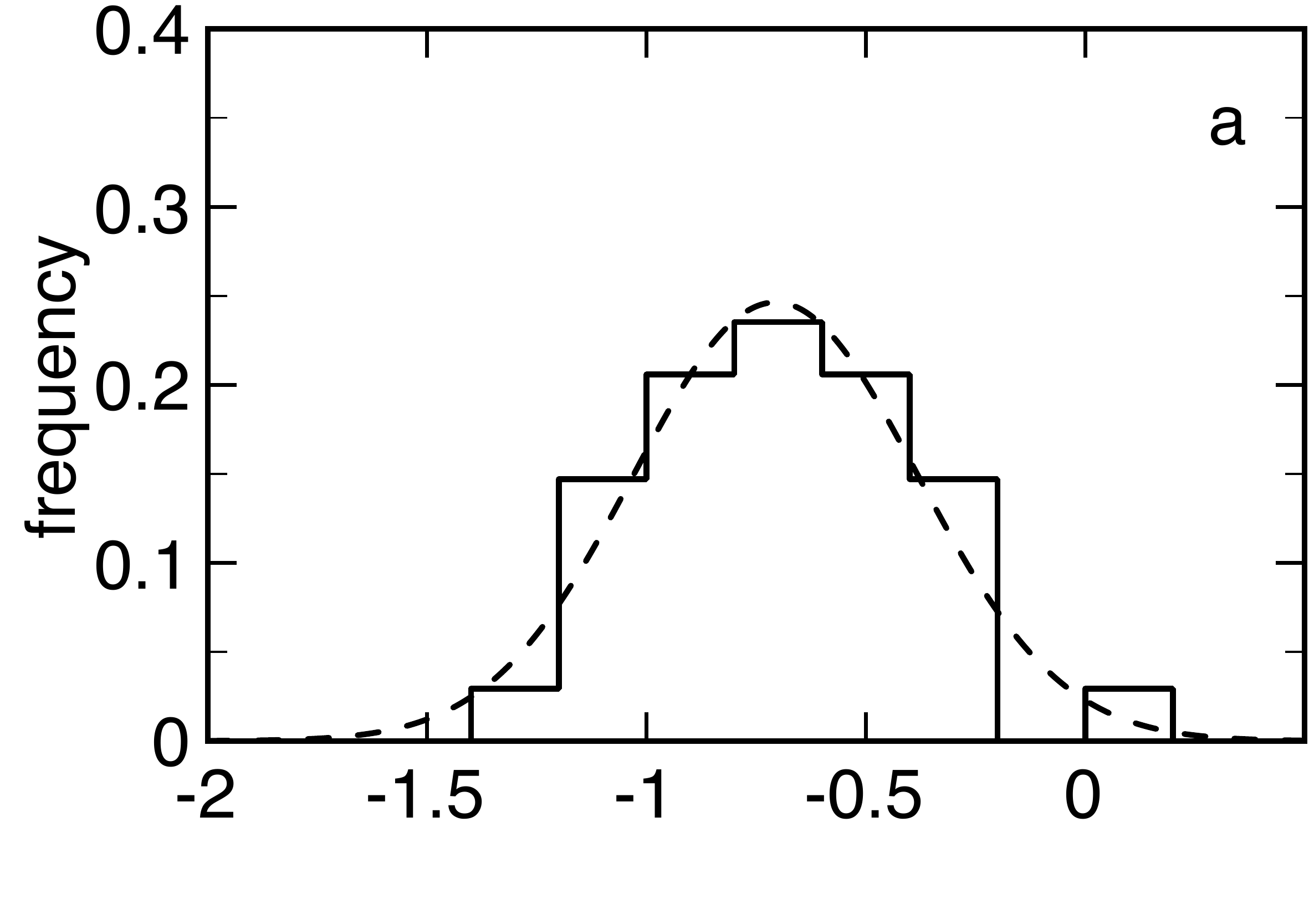}
\end{minipage}%
\begin{minipage}[c]{0.235\textwidth}
\includegraphics[width=1.0\textwidth]{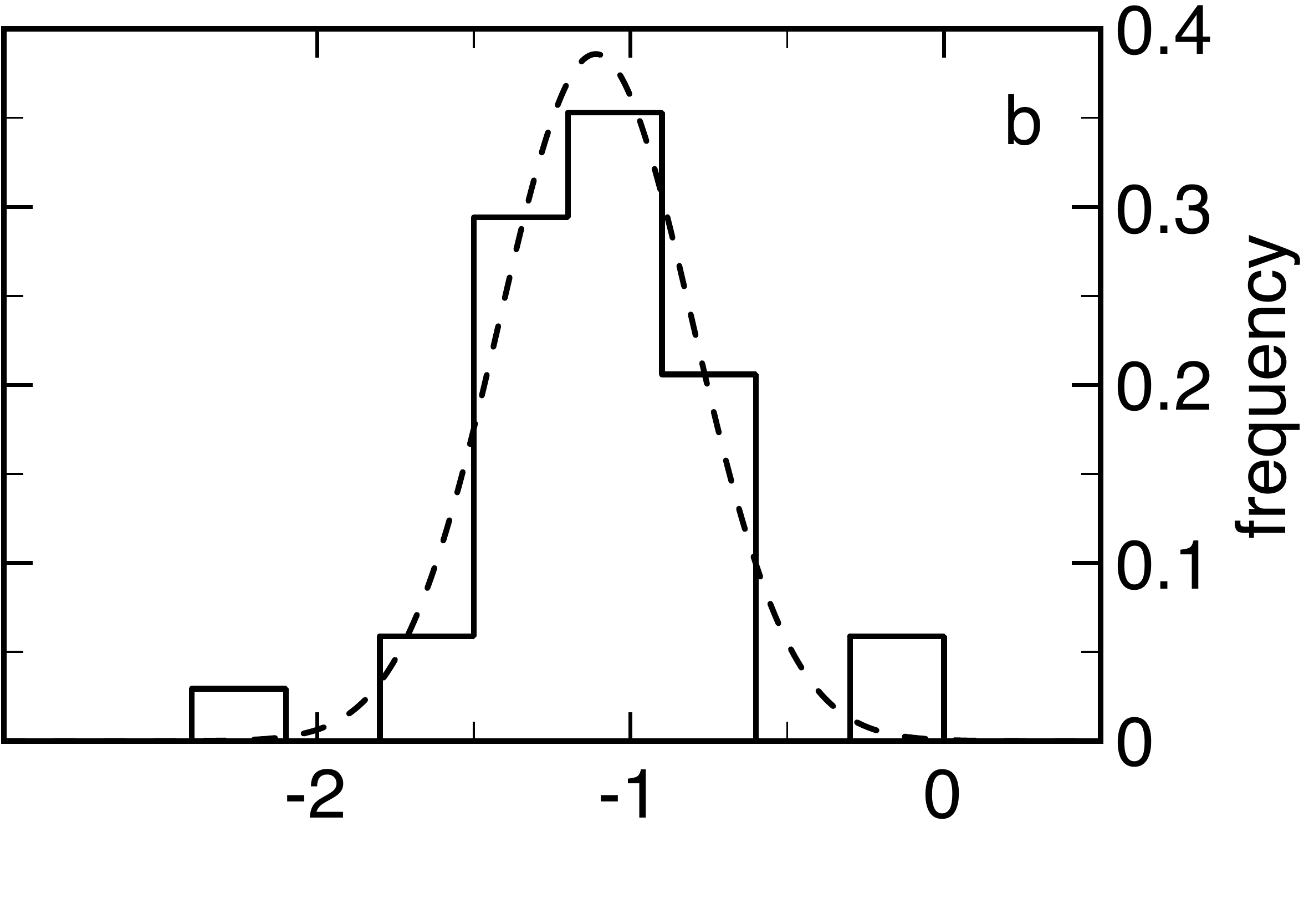}
\end{minipage}
\begin{minipage}[c]{0.235\textwidth}
\includegraphics[width=1.0\textwidth]{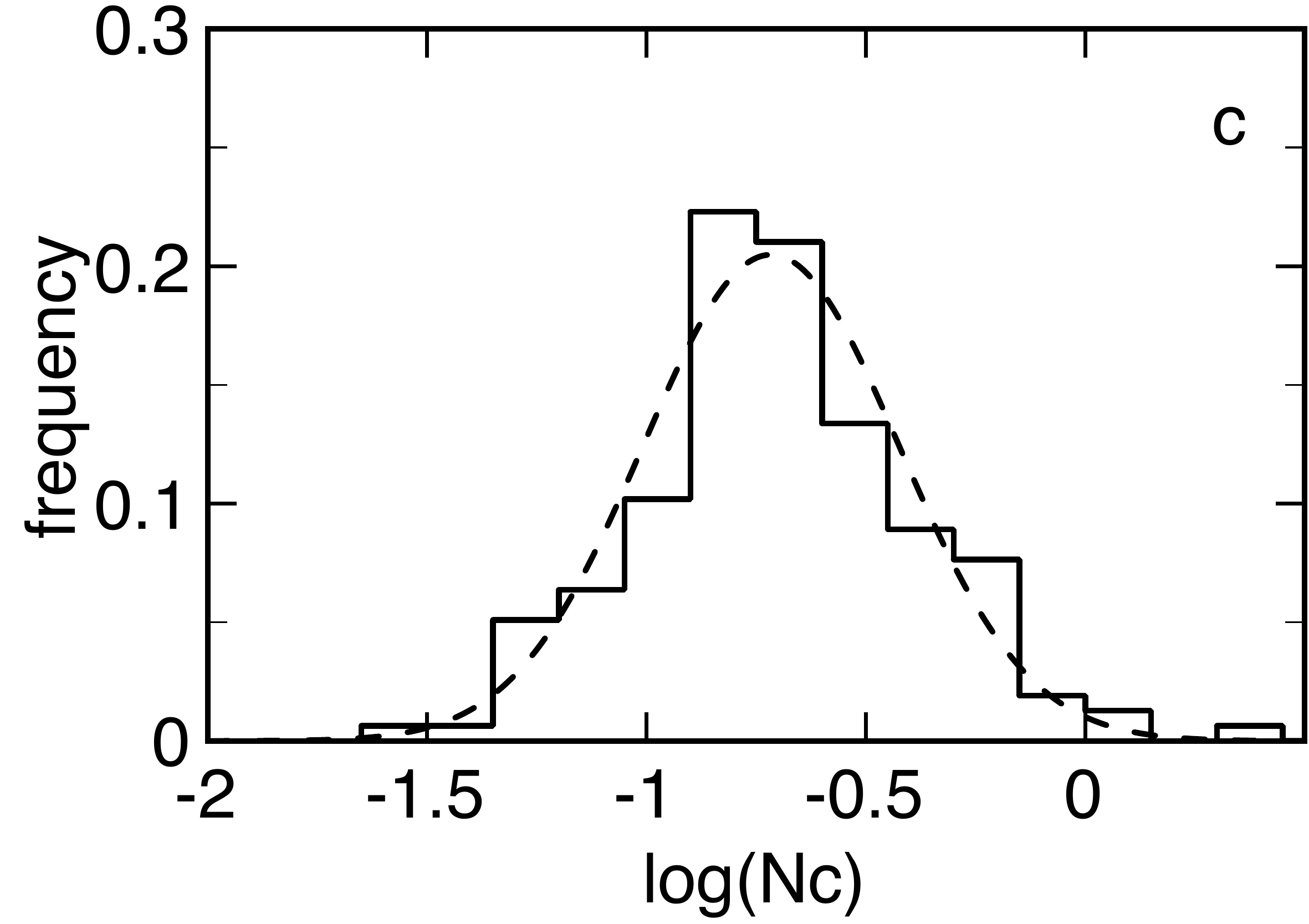}
\end{minipage}%
\begin{minipage}[c]{0.235\textwidth}
\includegraphics[width=1.0\textwidth]{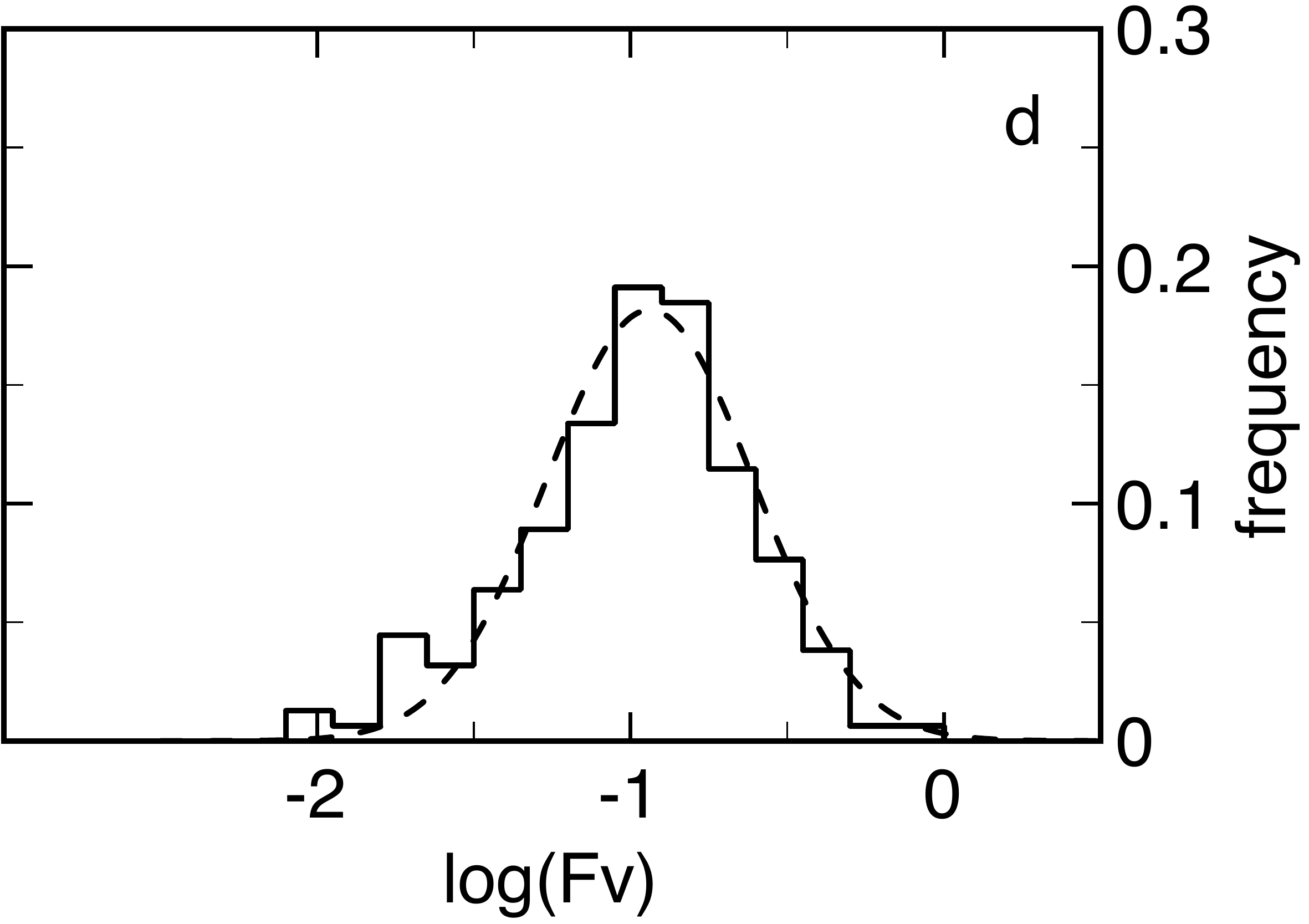}
\end{minipage}
\caption{Probability distribution functions of mean density in clouds, $\Nc$ (\textbf{a} and \textbf{c}) and volume filling factors, $\Fv$ (\textbf{b} and \textbf{d}). The dashed lines are the lognormal fits to the histograms; the fit parameters are given in Table~\ref{table:DIG}. Top: small sample (34 pulsars); bottom: BMM sample (157 pulsars).}
\label{fig:1}
\end{figure}

%% table
\begin{table*}
\caption{Lognormal fits to the PDFs of the DIG (Figs.~\ref{fig:1} and \ref{fig:2}).
The fitted function is $Y=(\sqrt{2\pi}\sigma)^{-1}\exp[-(\log_{10} X - \mu)^2/2\sigma^2]$.}
%\begin{center}
\begin{tabular}{llrrrr}
\hline
 & & \multicolumn{2}{c}{Position of maximum} & Dispersion & \\
Sample & $X$ & $\mu$ & $X$ & $\sigma$ & $\chi^2$  \\
\hline
This work & $\Nc$ ($\cmcube$) & $-0.71\pm 0.03$ 
	& $0.19\pm 0.02$ & $0.32\pm 0.03$ & $0.4$\\
$N=34$ & $\Fv$ & $-1.11\pm 0.03$
	& $0.078\pm 0.006$ & $0.31\pm 0.03$ & $0.8$\\
  & $\avg{\n}$ ($\cmcube$) & $-1.83\pm 0.02$ 
	& $0.015\pm0.001$ & $0.22\pm 0.02$ & $0.3$\\
  & $\avg{\n^2}$ ($\mathrm{cm}^{-6}$) & $-2.48\pm 0.02$ 
  	& $0.0033\pm 0.0002$ & $0.29\pm 0.02$ & $0.4$\\
  & & & & & \\
BMM  & $\Nc$ ($\cmcube$) & $-0.72\pm 0.02$ 
	& $0.19\pm 0.01$ & $0.29\pm 0.02$ & $1.7$\\
$N=157$ & $\Fv$ & $-0.94\pm 0.02$
	& $0.115\pm 0.005$ & $0.33\pm 0.02$ & $0.9$\\
  & $\avg{\n}$ ($\cmcube$) & $-1.68\pm 0.01$ 
	& $0.021\pm 0.001$ & $0.12\pm 0.01$ & $1.2$\\
  & $\avg{\n^2}$ ($\mathrm{cm}^{-6}$) & $-2.41\pm 0.03$ 
  	& $0.0039\pm 0.0003$ & $0.34\pm 0.03$ & $0.8$\\
\hline\noalign\\  
\end{tabular}
%\end{center}

\medskip
$Y$ is the fraction of sightlines in each bin divided by the logarithmic bin-width $\dif(\log_{10}X)$. $\chi^2$ is the reduced chi-squared goodness of fit parameter, with the error in each bin $\delta_i$ estimated as $\delta_i=\sqrt{N_i}$ and for (number of bins$-2$) degrees of freedom.
\label{table:DIG}
\end{table*}

In Fig.~\ref{fig:1}a we present the probability distribution function
(PDF) of the mean density in clouds, $\Nc$, for the sample of 34
pulsars. As the volume filling factor is (anti-) correlated with $\Nc$
(see BMM), we show the PDF of $\Fv$ for the same sample in Fig. 1b. In
log space, both PDFs are consistent with a Gaussian distribution, which
is equivalent to a lognormal distribution in linear space. The PDFs have
about the same dispersion, $\sigma$ (see Table~\ref{table:DIG}). The positions of the
maxima, $\mu$ = log(density of maximum), correspond to $\Nc = 0.19\pm 0.02 \cmcube$
and $\Fv = 0.078\pm 0.006$, which represent the centre of gravity in the
$\Fv$-$\Nc$ plot in fig.6 of \citet{Berkhuijsen:2008}.
This sample is rather small, with low counts $N_i$ in the histogram bins and (probably) overestimated Poisson errors $\delta_i=\sqrt{N_i}$ leading to rather small reduced-$\chi^2$ statistics (Table~\ref{table:DIG}). Therefore we also calculated the PDFs of
$\Nc$ and $\Fv$ for the much larger sample of BMM, which are shown in
Figs. 1c,d. Both are well fitted by Gaussians of widths that are nearly
identical to those of the small sample (see Table 1). The positions of
the maxima are at $\Nc = 0.19\pm 0.01 \cmcube$ and $\Fv = 0.115\pm
0.005$, corresponding to the centre of gravity in the $\Fv$-$\Nc$ plot
of BMM (their fig.11). The good agreement between the PDFs of the two
samples indicates that the statistical results on $\Fv$ and $\Nc$ of BMM
are not influenced by the model distances and statistical absorption
corrections that they used.

%% figure
\begin{figure}
\centering
\begin{minipage}[c]{0.235\textwidth}
\includegraphics[width=1.0\textwidth]{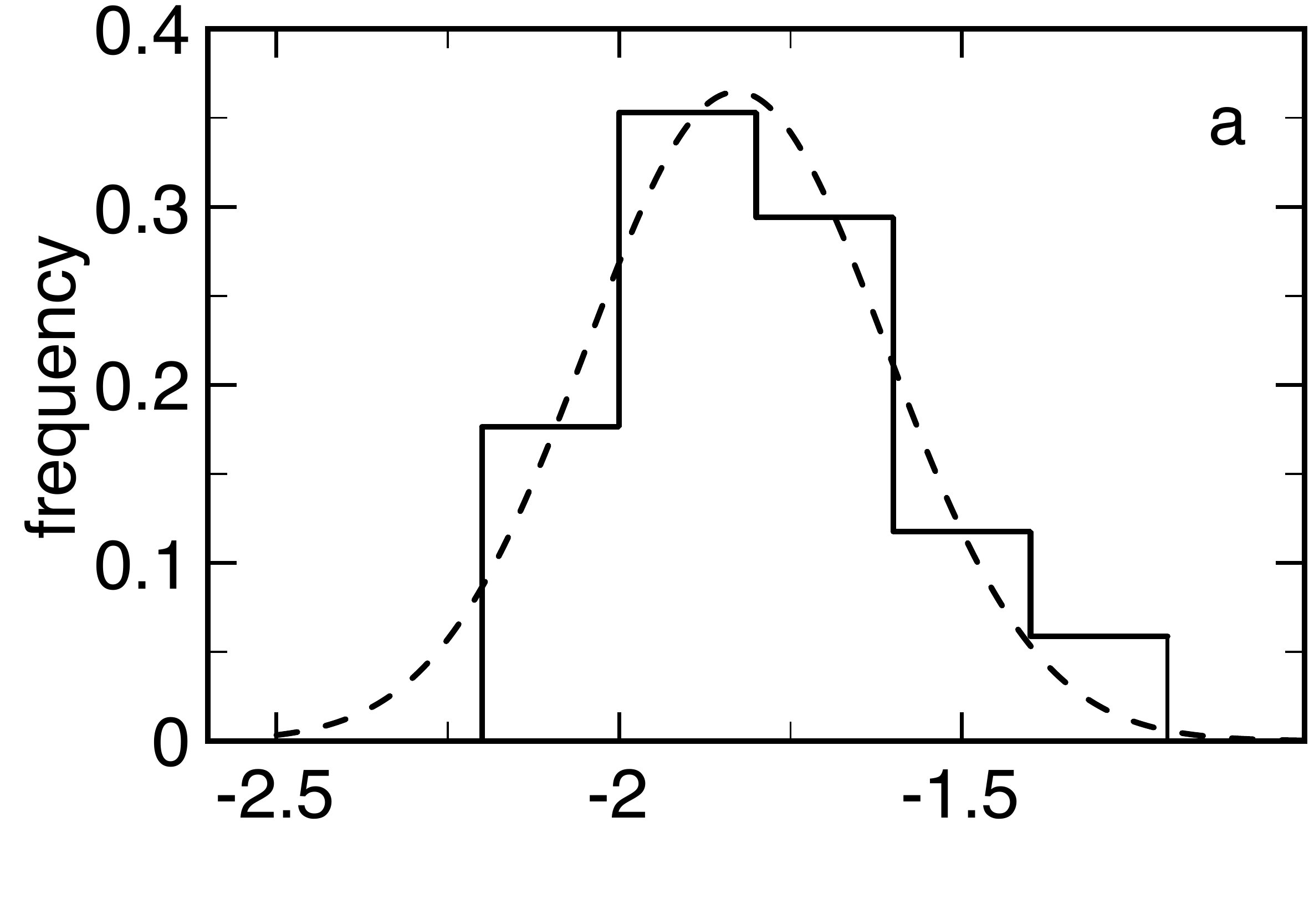}
\end{minipage}%
\begin{minipage}[c]{0.235\textwidth}
\includegraphics[width=1.0\textwidth]{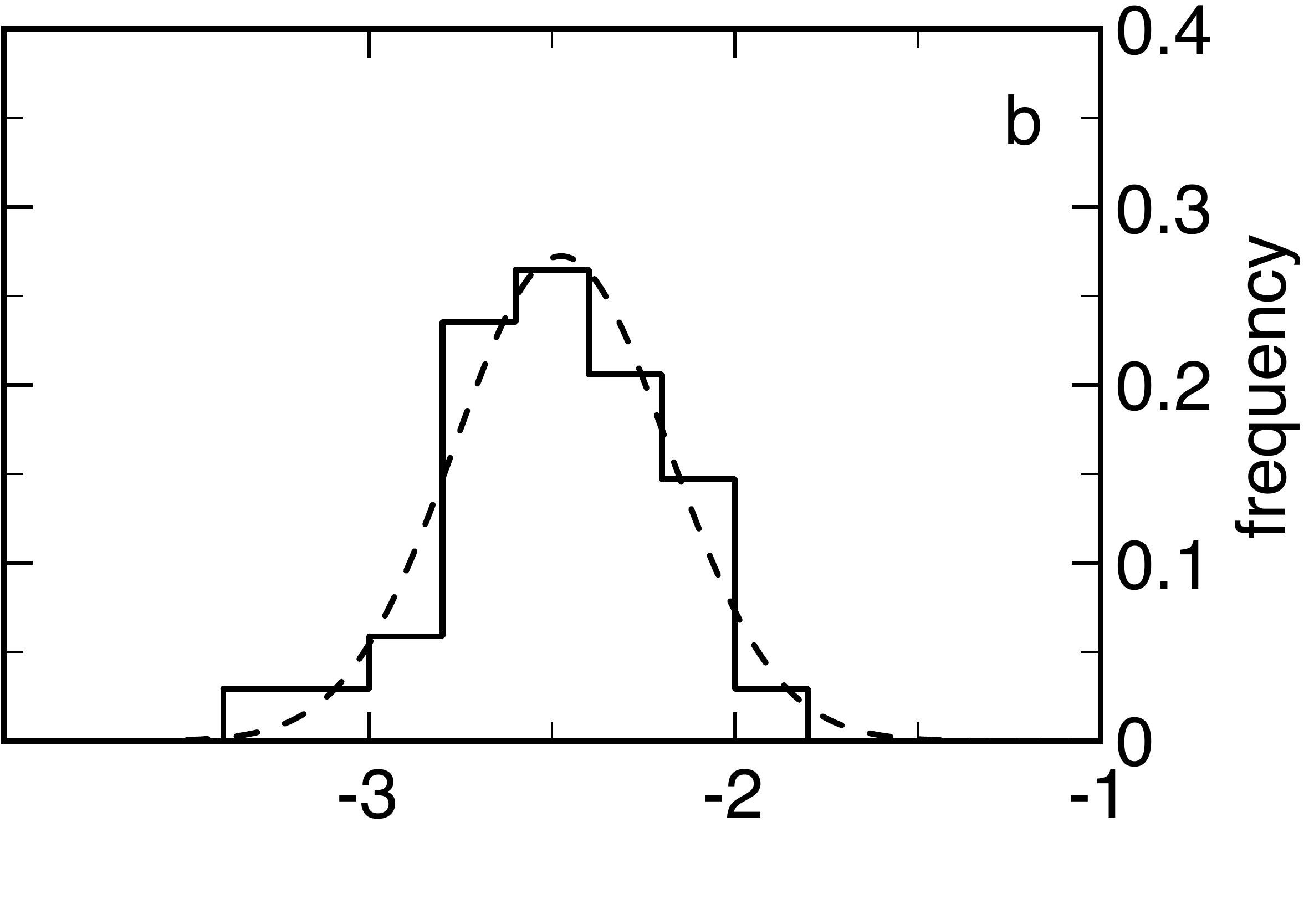}
\end{minipage}
\begin{minipage}[c]{0.235\textwidth}
\includegraphics[width=1.0\textwidth]{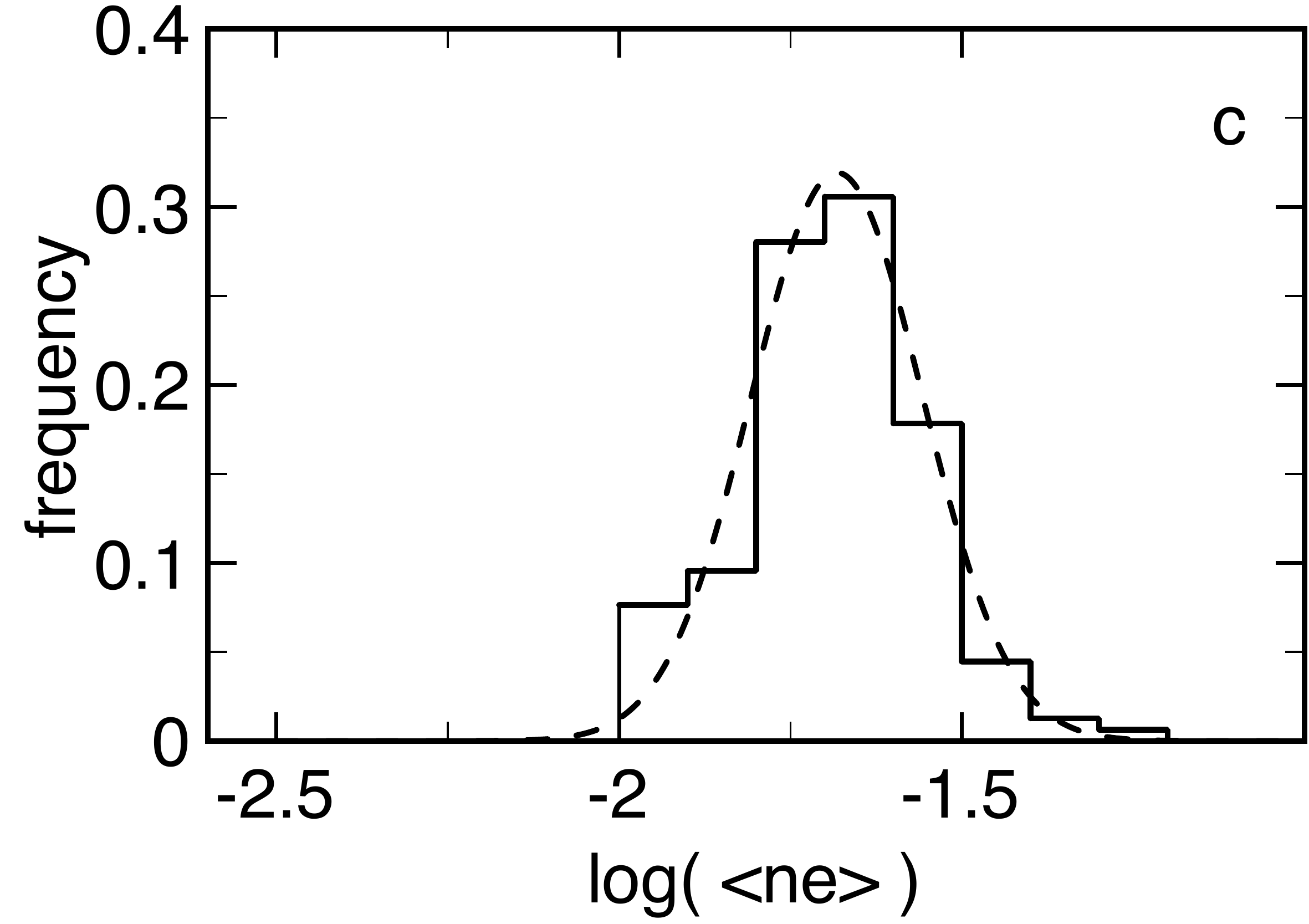}
\end{minipage}%
\begin{minipage}[c]{0.235\textwidth}
\includegraphics[width=1.0\textwidth]{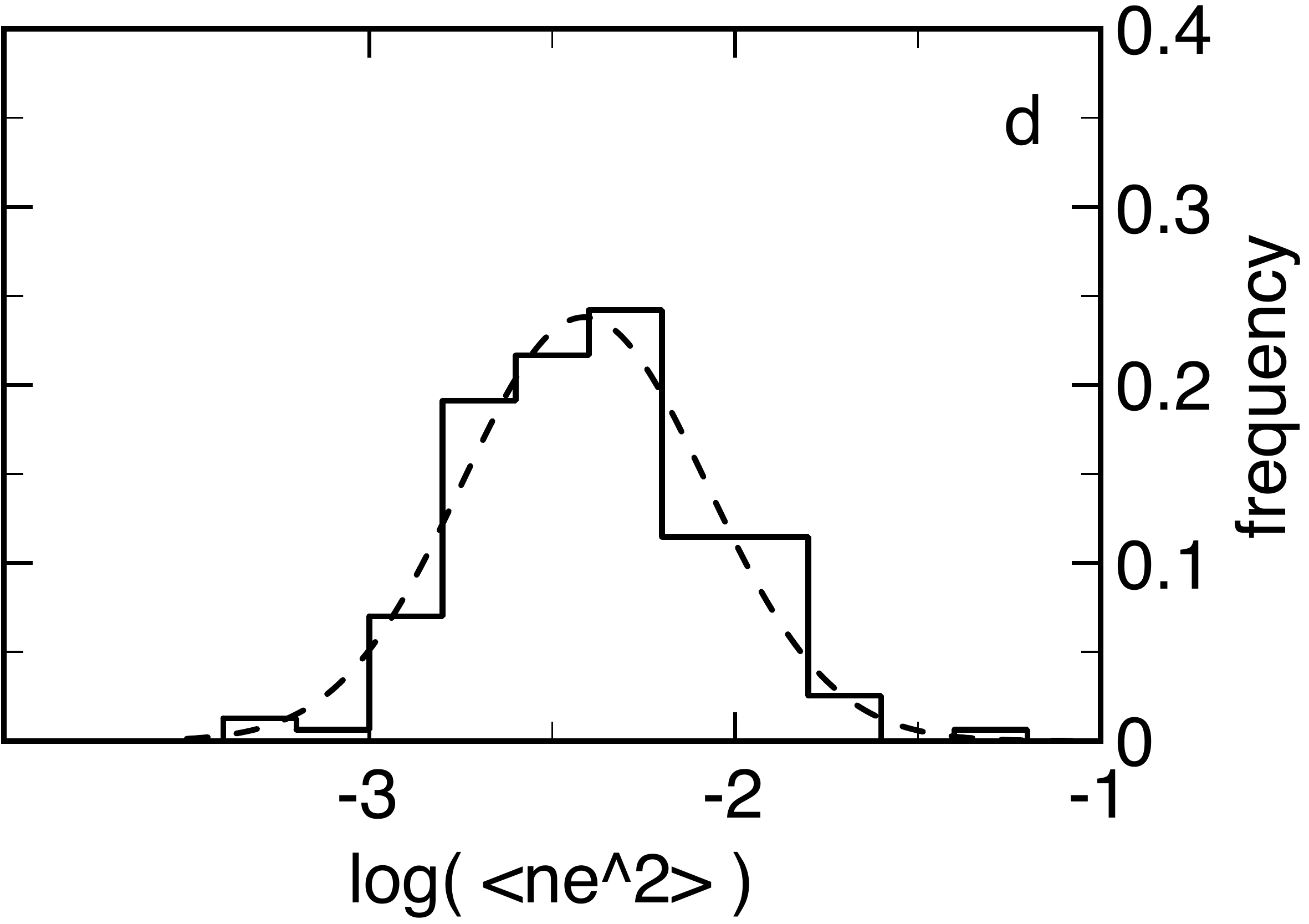}
\end{minipage}
\caption{ Probability distribution functions of $\avg{\n} = \DM/D$ (\textbf{a} and \textbf{c}) and $\avg{\n^2} = \EMp/D$ (\textbf{b} and \textbf{d}) for the small sample (top) and the BMM sample (bottom). The dashed lines are the lognormal fits to the histograms; the fit parameters are given in Table~\ref{table:DIG}.}
\label{fig:2}
\end{figure}

In Fig.~\ref{fig:2} we present the PDFs of the
average densities $\avg{\n}$ and $\avg{\n^2}$ for both samples, all of
which are well described by a lognormal distribution. The dispersion in
$\avg{\n}$ is smaller than the dispersions in $\Fv$ and $\Nc$ due to
their (anti-) correlation: $\Fv$ and $\Nc$ are not independent
random variables. Note that the dispersion of the PDF of  $\avg{\n}$ of the
BMM sample is about half that of the small sample. As BMM used distances
to the pulsars derived from the NE2001 model of \citet{Cordes:2002},
$\avg{\n}=\DM/D$ returns the densities of the model. The small dispersion
reflects the fact that the model is much smoother than the density
variations in the real ISM measured for the small sample. The dispersion
in $\avg{\n^2}$ is larger than that of $\avg{\n}$ as the intrinsic
spread in EM is much larger than in DM \citep[see BMM; ][]{Berkhuijsen:2008} while the distances used to calculate $\avg{\n^2}$ and
$\avg{\n}$ are the same.

It is interesting to compare our data with the results of the
magneto-hydrodynamic simulations of the ISM in the solar neighbourhood made by
\citet{Avillez:2005}. Their fig. 7 shows the density PDFs of five
temperature regimes that developed after about $400\Myr$. The curve for $8000 < T_e < 16000 \K$, which is most applicable to the
DIG, closely resembles a lognormal with maximum at $\log(n_{H}) = -0.75$
and dispersion $0.52$. The lognormal distribution extends over a much
larger density range (at least $-2.5 < \log(n_{H}) < 1.2$) than our
observations of $\Nc$ ($-2 < \log(\Nc) < 0$ in Fig. 1). The density of
the maximum of $0.18\cmcube$ agrees well with that of $\Nc = 0.19\pm
0.02\cmcube$ derived by us (see Table 1), but the dispersion is about
$70$ per cent larger. This could be due to the larger temperature range of this
component in \citet{Avillez:2005} compared to the $6000$-$10000 \K$
observed for the DIG \citep{Madsen:2006}. 

We conclude that the PDFs of the electron densities and filling
factors in the DIG in the solar neighbourhood are lognormal
as is expected for a turbulent ISM from numerical simulations.

\subsection{Density PDFs of diffuse $\HI$}
\label{subsec:results:HI}

%% figure
\begin{figure}
\centering
\begin{minipage}[c]{0.235\textwidth}
\includegraphics[width=1.0\textwidth]{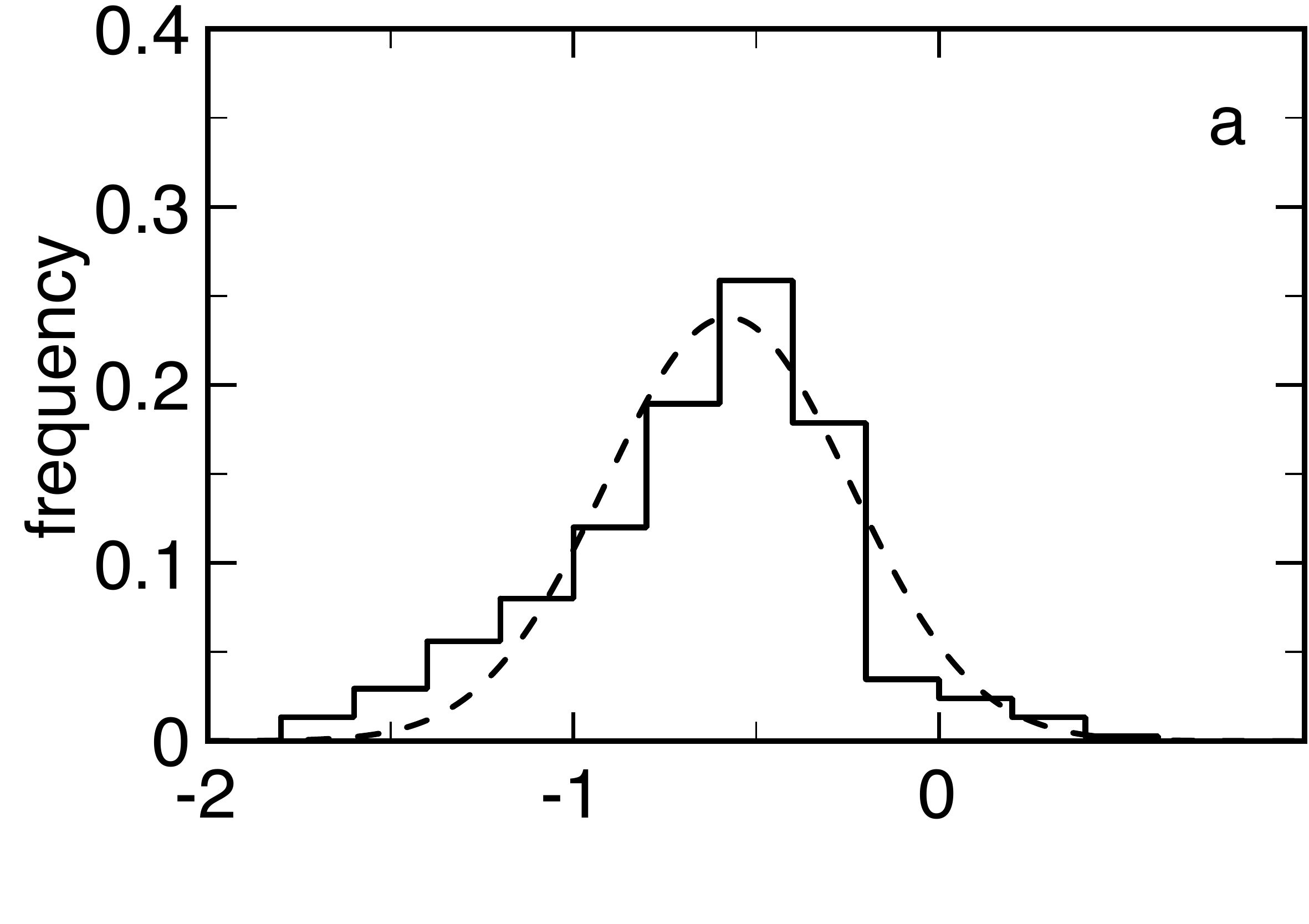}
\end{minipage}%
\begin{minipage}[c]{0.235\textwidth}
\includegraphics[width=1.0\textwidth]{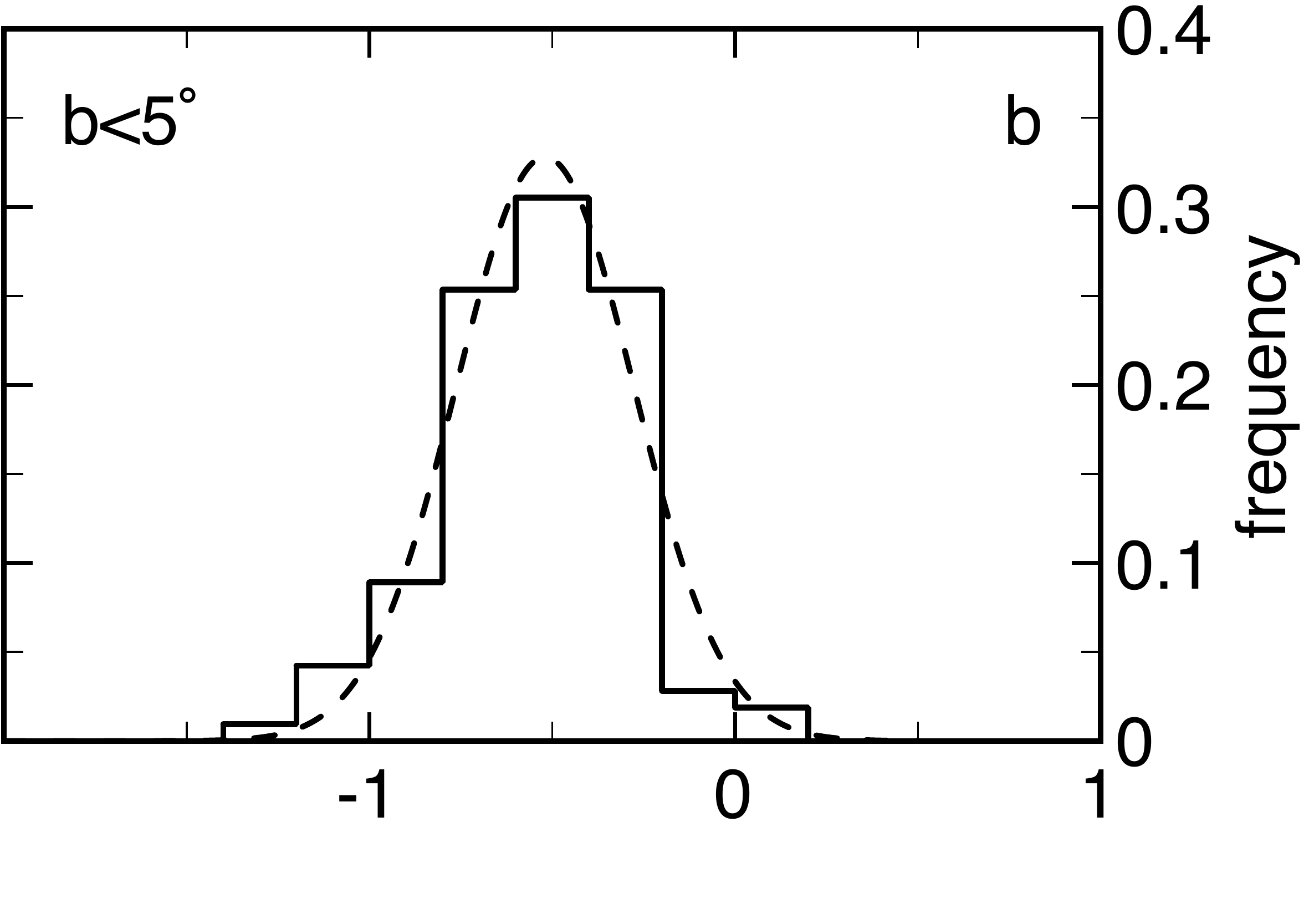}
\end{minipage}
\begin{minipage}[c]{0.235\textwidth}
\includegraphics[width=1.0\textwidth]{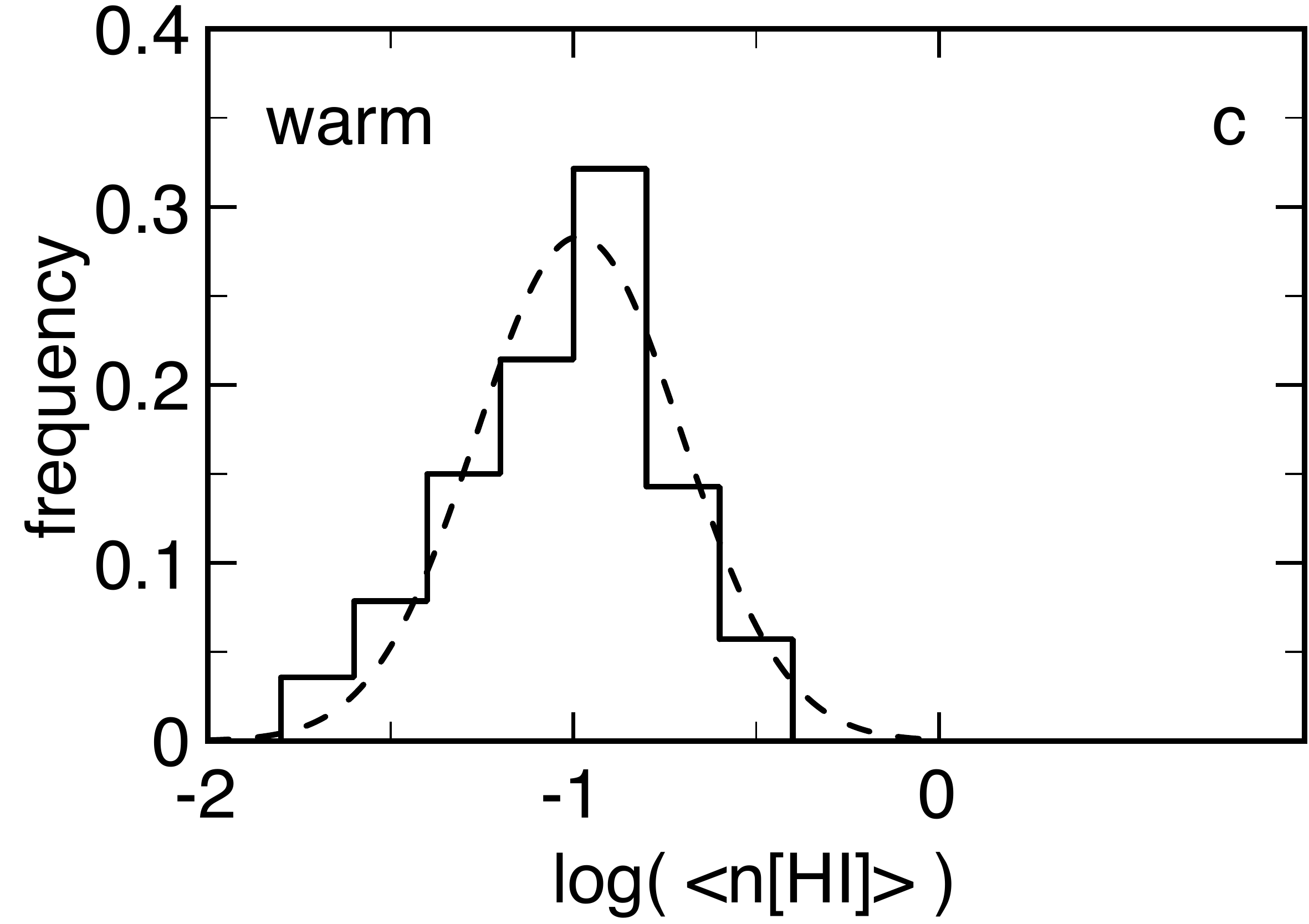}
\end{minipage}%
\begin{minipage}[c]{0.235\textwidth}
\includegraphics[width=1.0\textwidth]{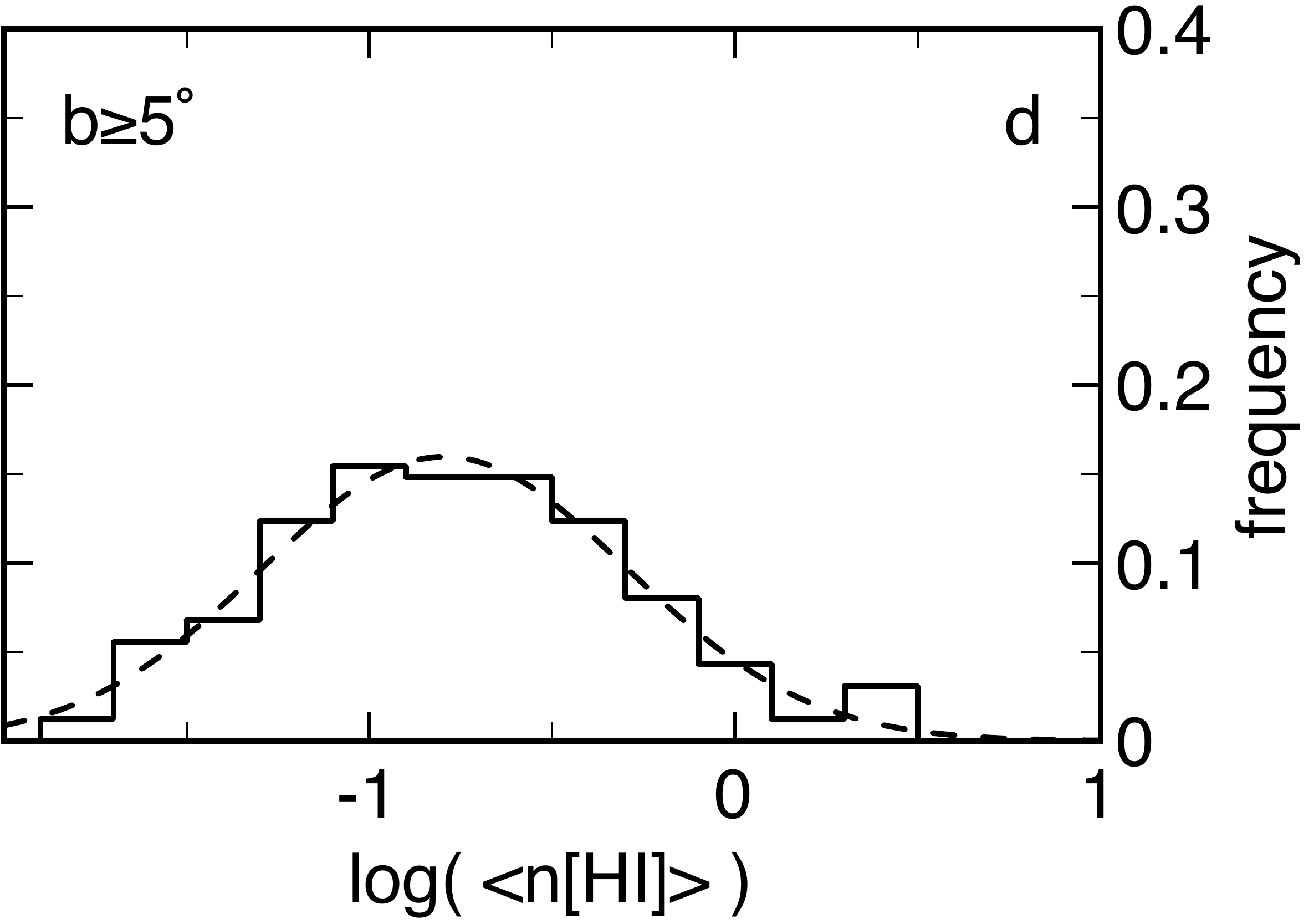}
\end{minipage}
\caption{Probability distribution functions of $\avg{\nH} = \NH/D$: \textbf{a}) Full sample of 375 stars \citep{Diplas:1994a}, of which are \textbf{b}) 213 stars at $|b|< 5\degree$ and \textbf{d}) 162 stars at $|b|\ge 5\degree$; \textbf{c}) 140 stars seen through warm $\HI$. The dashed lines are the lognormal fits to the histograms; the fit parameters are given in Table~\ref{table:HI}.}
\label{fig:3}
\end{figure}

%% table
\begin{table*}
\caption{Lognormal fits to the PDFs of $\avg{\nH}$ (Fig.~\ref{fig:3}). The fitted function is $Y=(\sqrt{2\pi}\sigma)^{-1}\exp[-(\log_{10} \avg{\nH} - \mu)^2/2\sigma^2]$.}
\begin{center}
\begin{tabular}{llrrrr}
\hline
 & & \multicolumn{2}{c}{Position of maximum} & Dispersion & \\
Area & $N$ & $\mu$ & $\avg{\nH}$ & $\sigma$ & $\chi^2$\\
\hline
All stars & $375$ & $-0.57\pm 0.03$ & $0.27\pm 0.02$ & $0.34\pm 0.03$ & $6.0$\\
  $|b|<5\degree$ & $213$ & $-0.52\pm0.02$ & $0.30\pm 0.02$ & $0.24\pm 0.02$ & $2.7$\\
  $|b|\ge 5\degree$ & $162$ & $-0.80\pm 0.02$ & $0.16\pm 0.01$ & $0.50\pm 0.02$ & $0.4$\\
  & & & & & \\
Warm $\HI$ & $140$ & $-0.98\pm 0.03$ & $0.10\pm 0.01$ & $0.28\pm 0.03$ & $2.5$\\
  $|b|<5\degree$ & $42$ & $-0.89\pm 0.01$ & $0.13\pm 0.01$ & $0.18\pm 0.01$ & $0.1$\\
  $|b|\ge 5\degree$ & $98$ & $-1.07\pm 0.03$ & $0.09\pm 0.01$ & $0.33\pm 0.03$ & $1.0$\\
\hline\noalign\\  
\end{tabular}
\end{center}

\medskip
$Y$ is the fraction of stars in each bin divided by the logarithmic bin-width $\dif(\log_{10}\avg{\nH})$. $\chi^2$ is the reduced chi-squared goodness of fit parameter, with the error in each bin $\delta_i$ estimated as $\delta_i=\sqrt{N_i}$ and for (number of bins$-2$) degrees of freedom.
\label{table:HI}
\end{table*}

In Fig.~\ref{fig:3}a we present the PDF of the average volume density of
$\HI$, $\avg{\nH}$, for the full sample of 375 stars of
\citet{Diplas:1994a}. Above $\mathrm{log}\avg{\nH} = -1$ the
distribution is approximately lognormal, but there is a clear excess at
lower densities reflected in the large reduced-$\chi^2$ statistic (see Table~\ref{table:HI}). Because low densities can be expected away from the Galactic
plane, we calculated the PDFs for the latitude ranges
$|b|<5\degree$ and $|b|\ge 5\degree$ separately, as shown 
in Figs.~\ref{fig:3}b and \ref{fig:3}d. Both distributions have a
lognormal shape but they are shifted with respect to each other: the
maximum of the low-$|b|$ sample is at $\avg{\nH}=0.30\pm0.02 \cmcube$ and that
of the high-$|b|$ sample at $\avg{\nH}=0.16\pm0.01 \cmcube$ (see Table 2). The
latter sample clearly causes the low-density excess in
Fig.~\ref{fig:3}a. 

The dispersion of the PDF of the high-$|b|$ sample is twice that of the low-$|b|$
sample. It is not clear whether this is a real difference or due to
selection effects in the low-$|b|$ sample. Stars at low Galactic
latitudes can only be seen through holes between the many dust clouds
and the low latitude sample may be biased towards low densities if the higher 
density diffuse gas ($\avg{\nH}\gtrsim 1\cmcube$) is associated with these clouds. 

\citet[ fig. 9]{Diplas:1994b} identified about 140 lines of sight probing
the warm diffuse $\HI$. This sample is especially interesting for
comparison with the DIG because the average gas temperatures of both components 
are $\sim8000$K. The PDF of the warm \HI\ (see Fig.~\ref{fig:3}c) is also
lognormal and peaks at $\avg{\nH}=0.10\pm0.01 \cmcube$ (see Table 2). All
densities are $<0.3 \cmcube$. Comparison with the full sample in
Fig.~\ref{fig:3}a shows that \emph{all} lines of sight with $\log(\avg{\nH}) <
-0.8$ (or $\avg{\nH} < 0.16 \cmcube$) probe the warm $\HI$. Table~\ref{table:HI} shows that the
dispersions of the PDFs of the warm $\HI$ are slightly smaller than for
the full sample. Clearly the combination of warm and cool (denser) gas in the full sample
increases the dispersion because the density range becomes larger.

%-----------------------------------------------------------------
\section{Discussion and Conclusions}

The results in Sect.~\ref{sec:results} show that the average volume densities of the 
DIG and the diffuse \HI\ within a few \kpc\ of the Sun follow a lognormal 
distribution, as is expected if the density is the result of a random, 
nonlinear process such as turbulence. 

The dispersions of the observed PDFs vary between about $0.2$ for the
DIG and for HI at $|b|<5\degree$, and $0.5$ for HI at $|b|\ge 5\degree$.
Can we understand such differences in the frame of the simulations?

The most remarkable difference is that the dispersions of the density
PDFs of HI at $|b|\ge 5\degree$ are about twice those at $|b|< 5\degree$
(see Table~\ref{table:HI}). If this is a real effect (see
Sect.~\ref{subsec:results:HI}), a possible explanation is that the low
latitude LOS cross more turbulent ``cells'', where the size
of a cell is related to the decorrelation scale of the turbulence, than
at high latitudes. Under the central limit theorem, one expects the
density PDF to become narrower as the number of cells along the
line-of-sight in a sample increases \citep{Vazquez-Semadeni:2001}. Since
the average distance to the stars is the same in both samples, this
would imply that the average size of a cell is smaller at low latitudes.
This would be consistent with the inverse dependence of the volume
filling factor \Fv\ on mean density in cells/clouds found for the DIG
\citep{Berkhuijsen:2006, Berkhuijsen:2008} and
diffuse dust \citep{Gaustad:1993}. Higher density means smaller \Fv\,
hence smaller clouds (as fewer clouds at low $|b|$ is unlikely).
\citet{Vazquez-Semadeni:2001} used models of isothermal turbulence to
investigate the relation between the shape of average density PDFs and
the number of turbulent cells along the line-of-sight. Their results
suggest that the dispersion in density will narrow by a factor of $\sim
2$ if the number of cells increases by a factor of $\sim 5$.

Another interesting difference exists between the dispersions of the
sample of warm HI at $|b|\ge 5\degree$ and that of $\avg{\n}$ of the DIG (small sample),
which is at the same latitudes. The temperatures of the two components are
similar and if the ionized and atomic gas are well mixed, one would expect
their dispersions to be the same. However, the dispersion of the DIG sample,
$0.22\pm 0.01$, is about $30$ per cent smaller than that of the warm \HI\ sample, $0.33\pm0.03$ 
(see Tables~\ref{table:DIG} and \ref{table:HI}). A plausible explanation for the difference, which is also consistent with the higher density of the maximum in the diffuse $\avg{\nH}$ PDF, is that low density regions are more readily ionized than higher density gas and that the average degree of ionization of the diffuse gas is substantially lower than $50$ per cent. We estimate the degree of ionization to be about $14$ per cent, using the densities for $\avg{\n}$ (small sample) and $\avg{\nH}$ in Tables~\ref{table:DIG} and \ref{table:HI}, consistent with the results of \citet[ their fig.~13]{Berkhuijsen:2006} for a mean height above the mid-plane of about $500\pc$. Alternatively, the DIG could have a higher mean temperature than the warm \HI\ but with a smaller temperature range; in the simulations of \citet{Avillez:2005} high temperature gas indeed has a lower median density and smaller dispersion.

Several groups have noted a link between the rms-Mach number $\mathcal{M}$ and the dispersion of the gas density PDF in isothermal numerical simulations \citep[e.g.][]{Padoan:1997, Passot:1998, Ostriker:2001}. Although the DIG is not isothermal \citep{Madsen:2006}, to a first approximation it can be considered so because the sound speed scales as $c_s\sim T^{1/2}$ and the observed temperature range of $6000\K<T<10000\K$ corresponds to only a 30\% difference in $c_s$. Then using the formula $\sigma_{\ln}^2=\ln(1+\beta^2\mathcal{M}^2)$ with $\beta\approx 0.5$ given by \citet{Padoan:1997}, and a typical value of $\sigma\simeq 0.3=\sigma_{\ln}/\ln(10)$ from our results for the DIG and warm \HI\, we obtain $\mathcal{M}\simeq 1.6$. \citet{Hill:2008} compared their observed Emission Measure PDFs for the DIG with PDFs derived from isothermal MHD turbulence simulations to find reasonable agreement for $1.4<\mathcal{M}<2.4$, consistent with our estimated value. A typical DIG temperature of $T\simeq 8000\K$ gives $c_s\simeq 12.5\kms$ and $\mathcal{M}\simeq 1.6$ would then require turbulent velocities of $v\simeq 20\kms$.

While the global-disc simulations of \citet{Wada:2007} produced lognormal 
density PDFs, their dispersions are about 4 times greater than the 
dispersions we have found. This may be related to the  
average gas densities in their simulations:  $\avg{\nH}\sim1\cmcube$ compared to 
e.g. $\avg{\nH}\sim 0.3\cmcube$ for our total \HI\ sample. 

We may draw the following conclusions from the discussion of our results:
\begin{enumerate}
\item The density PDF of the diffuse ISM cannot be fitted by one lognormal.
\item The density PDFs of the diffuse ISM in the disk ($|b|<5\degree$) and away from the disk ($|b|\ge 5\degree$) are lognormal, but the positions of their maxima and the dispersions differ.
\item Several effects seem to influence the shape of the PDF. An increase of the number of clouds/cells along the LOS causes a \emph{decrease} in the dispersion and a shift of the maximum to higher densities. On the other, hand, an increase in the average density (or decrease in the mean temperature) \emph{increases} the dispersion as well as the density of the maximum.
\end{enumerate}

The competing effects described in the last conclusion will complicate the
interpretation of PDFs observed for external galaxies.

\citet{Elmegreen:2002} and \citet{Wada:2007} have shown that the star formation 
rate in a galaxy is related to the shape of the density PDF if it is lognormal:
the dispersion is an important parameter in this respect \citep{Tassis:2007, Elmegreen:2008}. A better
understanding of the factors that influence the dispersion of the lognormal
density PDF may be obtained from future simulations.

%__________________________________________________________________________
\section{Summary}

Lognormal density PDFs have been found 
in recent numerical simulations of the ISM -- both local, isothermal models 
\citep[e.g.][]{Vazquez-Semadeni:2001, Ostriker:2001, Kowal:2007} and multi-phase, 
global models \citep{Avillez:2005, Wada:2007} -- and have become an important component of 
theories of star formation \citep{Elmegreen:2002, Tassis:2007, Elmegreen:2008}. To date there has been little 
observational data with which to compare density distributions produced by the 
simulations. 

The results reported here provide strong support for the existence 
of a lognormal density PDF in the diffuse (i.e. average densities of 
$n<1\cmcube$) ionized and neutral components of the ISM. In turn, the form of 
the PDFs is consistent with the small-scale structure of the diffuse ISM being controlled 
by turbulence. Future simulations should allow the calibration of the dispersion of the 
diffuse gas density PDF in terms of physically interesting parameters, such as 
the number of turbulent cells along the line of sight.

%__________________________________________________________________________
\section*{Acknowledgments}
We thank Dr. Rainer Beck for comments on an earlier version of the manuscript, Dr. Brigitta von Rekowski for advice on fitting PDFs and the referee for helpful suggestions on data presentation and interpretation. AF thanks the Leverhulme Trust for financial support under research grant
F/00~125/N.

%__________________________________________________________________________

\bsp

\label{lastpage}

\end{document}